# Self-adjoint extension procedure for a singular oscillator

Anzor Khelashvili[1] and Teimuraz Nadareishvili[1,2]

[1] Inst. Of High Energy Physics, Iv. Javakhishili Tbilisi State University, Tbilisi, Georgia.

[2] Depatment of Physics, Faculty of Exact and Natural Sciences, Iv. Javakhishili Tbilisi State University,

Tbilisi, Georgia.

**Abstract**. For a singular oscillator, the Schrödinger equation is obtained an equation of eigenvalues, and the dependence of energy on the self-adjoint extension parameter is established. It is shown that the self-adjoint extension violates the well-known property of equidistance of energy levels for the oscillatory potential, well-known in quantum mechanics. The concept of quantum defect is generally introduced, and the wave function of the problem is written as a single function.

# I . Introduction

Recently much attention has been devoted to the problems of self –adjoint extension (SAE) for the inverse square behaved potentials in the Schrödinger equation Number of physically significant quantum-mechanical problems manifest in such a behavior. Examples of such systems are: Valence electron model for hydrogen like atoms in Scrodinger euation [2], Coulomb and Hulthen problems in the Klein-Gordon and Dirac equations [3], the theory of black holes [4], conformal quantum mechanics [5], Aharonov-Bohm effect [6], Dirac monopoles [7], quantum Hall effect [8], Calogero model [9] and etc.

Detailed consideration of above-mentioned problems puts in doubt the motivations for neglecting of so-called additional (singular) solutions, which are based on mathematical sets of quantum mechanics without invoking of specific physical ideas.
Motives will be given below, according to which additional solutions arising due to singularity are sometimes avoided. We have shown in the work [10] that neither motivation is completely argumentative and it is necessary to maintain additional solutions, which in turn led to the introduction of the self-adjoint expansion procedure for the bound states. In this paper, the singular solution problem is discussed, that is, the self-adjoint expansion procedure for the singular oscillator.

The paper is oganized as follows: Chapter II briefly analyses the content of the problem in the Schrödinger equation in the case of a discrete spectrum and shows that it is necessary to maintain additional solutions under certain conditions. Chapter III introduces the self-adjoint extension parameter. In Chapter IV, the self-adjoint extension procedure for the singular oscillator is carried out. The main results of the paper are briefly summarized in the conclusion.

## II. Content of the problem in the Schrödinger equation

Based on various physical requirements, we have shown in [10-12] that the complete radial function in the origin should have the following behaviour

$$\lim_{r \to 0} rR = 0 \tag{2.1}$$

In the Schrödinger equation, one usually sees regular potentials that at the oiginsatisfy the condition

$$\lim_{r \to 0} r^2 V = 0 \tag{2.2}$$

Then the radial wave function at the origin behaves like this [13,14]

$$R_{r \to 0} = C_1 r^l + C_2 r^{-(l+1)} \tag{2.3}$$

where $l$ - is an orbital moment. It is obvious that the second term in this representation is singular at the origin - it does not satisfy the condition (2.1) and therefore it should be neglected ($C_2 = 0$).

It is also known that for the so-called singular potentials whose behaviour is

$$\lim_{r \to 0} r^2 V \to \pm\infty \tag{2.4}$$

We have a case of "falling" on the center [15-18].

It is interesting to study the intermediate (transitional) behavioral potentials

$$\lim_{r \to 0} r^2 V \to \pm V_0 \quad (V_0 = const > 0) \tag{2.5}$$

In (2.5) the two signs correspond to repulsion (+) and attraction (-).

It is necessary to note that in the work [10] we showed that for the potentials of type (2.4) and (2.5) due to the singularity of the Laplacian operator in the origin, after inserting $R(r) = \dfrac{u(r)}{r}$, we do not get a standard equation for the function $u(r)$, but an additional term containing the delta function appears in the equation [11,12] and to remove it the function $u(r)$ must have the following behavior at the beginning $u(r) \underset{r \to 0}{\approx} r^{1+v}$, where $v$ is zero or a positive integer according to the theory of generalized functions. Such behavior takes place only for regular (2.2) potentials. As for the singular (2.4) and (2.5) potentials, we must work with full radial functions for them.

In the case of attraction, we can formulate the following theorem [10]:

**Theorem.** Schrödinger's equation, in addition to standard solutions for potentials of type (2.5), necessarily has additional solutions if the condition is fulfilled

$$l(l+1) < 2mV_0 \tag{2.6}$$

Indeed (2.5) for the attraction potentials, the Schrödinger equation at short distances looks like this

$$R'' + \frac{2}{r}R' - \frac{P^2 - 1/4}{r^2}R = 0 \tag{2.7}$$

where

$$P = \sqrt{(l + 1/2)^2 - 2mV_0} > 0 \tag{2.8}$$

and equation (2.7) has such a solutionၹა

$$\lim_{r \to 0} R = a_{st} r^{-1/2+P} + a_{add} r^{-1/2-P} = R_{st} + R_{add} \tag{2.9}$$

Thus we have two intervals for the parameter. in the interval

$$0 < P < 1/2 \tag{2.10}$$

We have to keep the second term $a_{add} r^{-1/2-P} = R_{add}$, because the boundary condition (2.1) is fulfilled for it. The potential of type (2.5) was first discussed by K. Case [15], but he neglected the second term in the solution. As for $P \geq \frac{1}{2}$, we should keep only the first member $a_{st} r^{-1/2+P} = R_{st}$

From the relations (2.8) and (2.10) the condition (2.6) for the presence of additional vehicles is derived. If we require that $P$ is real number(otherwise we have the so-called "falling" event on the center [15-18]), the parameter should be limited by the following condition

$$2mV_0 < l(l+1) + 1/4 \tag{2.11}$$

The last two inequalities convert the magnitude $2mV_0$ into the following interval

$$l(l+1) < 2mV_0 < l(l+1) + 1/4 \tag{2.12}$$

Thus from (2.6) we see that in the $l = 0$ states except the standard solutions there are additional solutions as well for arbitrary small $V_0$, while for $l \neq 0$ states the "strong" field is necessary in order to fulfill (2.6).

It should be mentioned, that additional solutions survive such traditional requirement as the normalizability of wave function [18,19] and the integral from probability density is finite [20]. More stronger requirements for the wave function are discussed in the manuals [21, 22]. In particular, it is required that the matrix elements of the kinetic energy operator are finite. In [10]we have shown that this is too strict and physically unjustified a requirement, and that the total energy must be finite! This condition satisfy additional solutions!

To summarize all above-mentioned restrictions and comments as well as other artificial ones, we conclude that there is no satisfactory requirement in the framework of quantum mechanics, which avoids this additional solution self-consistently.Therefore, one has to retain this additional solution and study its consequences.

## Ⅲ. **Introducing the self-adjoint extension parameter**

It is known that for attraction potentials of the type (2.4) and (2.5) in the Schrödinger equation, for the so-called case of "falling" on the center, it is not enough to know only the potential and it is necessary to introduce any constant [15-18]. Note that, as can be seen from the definition (2.8), this corresponds to the case when

$$2mV_0 > (l + 1/2)^2 \tag{3.1}$$

In mathematical language, this means that the Hamiltonian of the problem is symmetric (Hermitian), but is not a self-adjoint operator, its defect index is (1,1), and it is necessary to introduce one parameter to make the Hamiltonian a self-adjoint operator [23-24]. Mathematical self-adjoint expansion is a rather difficult and time-consuming operation [23-24]. Therefore, it is more convenient to use the alternative and fast so-called "pragmatic approach" [25], which gives the same results as the self-adjoint extension. In particular, it is shown in the work [25] that in this approach the energy eigenfunctions form a complete orthonormal set, and the energy values are real! (These properties the self-adjoint Hamiltonian operator). But this is a non-physical case, because the particle "falls" to the centre, and so its energy is ubounded from below.

As for area

$$2mV_0 < (l+1/2)^2 \qquad (3.2)$$

As we mentioned above, in the interval (2.10) it is necessary to maintain additional solutions, and in this case it is easy to show that for any two energy eigenfunctions $u_1$ and $u_2$, which correspond to the $E_1$ and $E_2$ energies, the orthogonality condition is as follows [10]

$$m(E_1 - E_2)\int_0^\infty R_2 R_1 r^2 dr = P\{a_1^{st} a_2^{add} - a_2^{st} a_1^{add}\} \qquad (3.3)$$

where $a_i^{st}$ and $a_i^{add}$ ($i=1,2$) are constants defined in the ratio (2.9).

Naturally, the question arises how to satisfy the condition of orthogonality? It is clear that we must demand the fulfillment of the following condition

$$a_1^{st} a_2^{add} - a_2^{st} a_1^{add} = 0 \qquad (3.4)$$

i.e.

$$\frac{a_{1\,add}}{a_{1\,st}} = \frac{a_{2\,add}}{a_{2\,st}} \qquad (3.5)$$

In this case, the radial $\hat{H}_R$ Hamiltonian operator becomes a self-adjoint operator. This generalizes the result of Case [15], where only standard solutions are considered.

Thus, it is necessary to introduce the so-called the self-adjoint extension parameter, which in our case is defined as follows

$$\ddagger_B \equiv \frac{a_{add}}{a_{st}} \qquad (3.6)$$

$\ddagger_B$ the parameter is also the same for all levels (fixed $l$ orbital moment) and it is real for bound states.

## IV. Singular Oscillator

Let's use the formalism developed above, for the (2.5) type singular oscillator problem, i.e. when the potential has the following form

$$V = -\frac{V_0}{r^2} + gr^2;\ V_0, g > 0 \qquad (4.1)$$

It should be noted that the potential of many physical problems [4,8,9] goes on the potential of type (4.1).

(4.1) for the potential, the Schrödinger equation for the radial function takes the following form [18]

$$\langle R'' + \frac{3}{2} R' + \left[ n + s + \frac{3}{4} - \frac{\langle}{4} - \frac{s(s+1/2)}{\langle} \right] R = 0 \qquad (4.2)$$

where
$$\varsigma = \sqrt{2mg}\, r^2 \qquad (4.3)$$
$$l(l+1) - 2mV_0 = 2s(s+1) \qquad (4.4)$$
$$\sqrt{\frac{2m}{g}}\, E = 4(n+s) + 3 \qquad (4.5)$$

The solution of (4.2) for large $\varsigma$ - behaveses as $e^{-\frac{\varsigma}{2}}$, and for small $\varsigma$ - as $\varsigma^s$, where $s$ is defined as the positive root of the equation (4.4), which can be (by using (2.8) notations) written as (2.8)

$$s = \frac{1}{4}\left[-1 + \sqrt{(2l+1)^2 - 8mV_0}\right] = \frac{1}{2}\left[-\frac{1}{2} + P\right] \qquad (4.6)$$

Therefore, we are looking for a solution in the following way

$$R = e^{-\frac{\varsigma}{2}} \varsigma^s w \qquad (4.7)$$

and for the $w$ function we get the equation for the degenerate hypergeometric function

$$\varsigma w'' + \left(2s + \frac{3}{2} - \varsigma\right) w' + n w = 0 \qquad (4.8)$$

This equation has four independent solutions, any two of which determine the fundamental system of solutions [26]. They are (in the notation of [26]):

$$\begin{aligned} y_1 &= F(a,b;\varsigma) \\ y_2 &= \varsigma^{1-b} F(1+a-b, 2-b; \varsigma) \\ y_5 &= \Psi(a,b;\varsigma) \\ y_7 &= e^{\varsigma} \Psi(b-a, b; -\varsigma) \end{aligned} \qquad (4.9)$$

where

$$a = -n, \quad b = 2s + \frac{3}{2} \qquad (4.10)$$

As a rule, only $y_1$ is discussed in scientific articles and textbooks (see for example [2,18]). For example $a = -n$ demand gives standard levels. Other solutions ($y_2, y_5, y_7$) have a singular behavior at the origin, which is why they are often neglected, but as we mentioned above, the singularities, for attraction type (2.5) potentials, are of the $r^{-\frac{1}{2}-P}$ type and in the $0 < P < 1/2$ area should be considered.

Let's first consider $y_1$ and $y_2$ pair. In this case, (4.2)-s general solution will be

$$R = e^{-\frac{\sqrt{2mg}}{2} r^2} \left\{ C(2mg)^{\frac{P-\frac{1}{2}}{4}} r^{\frac{1}{2}+P} F\left(-n, 1+P; \sqrt{2mg}\, r^2\right) + \right. \\ \left. + D(2mg)^{\frac{-P-\frac{1}{2}}{4}} r^{\frac{1}{2}-P} F\left(-n-P, 1-P; \sqrt{2mg}\, r^2\right) \right\} \qquad (4.11)$$

From the behaviour of (4.11) at the origin and from (3.5), (3.6) we get the parameter of self-adjoint extension

$$\ddagger = \frac{D}{C} \tag{4.12}$$

On the other hand, note that R must fall at infinity and from this requirement and the well-known behaviour of confluent hypergeometric functions F at infinity [27], we obtain the following equation

$$C_1 \frac{\Gamma(1+P)}{\Gamma(-n)} + D \frac{\Gamma(1-P)}{\Gamma(-n-P)} = 0 \tag{4.13}$$

from which using (4.5), (4.6) and (4.12) we get the eigenvalue equation

$$\frac{\Gamma\left(-\frac{1}{4}\sqrt{\frac{2m}{g}}E + \frac{1}{2} - \frac{P}{2}\right)}{\Gamma\left(-\frac{1}{4}\sqrt{\frac{2m}{g}}E + \frac{1}{2} + \frac{P}{2}\right)} = -\ddagger \frac{\Gamma(1-P)}{\Gamma(1+P)} \tag{4.14}$$

As we can see (4.14) is a complex transcendental equation for the energy E depending on the parameter τ and only for three values of the parameter τ we obtain the analytical solution of this equation:
i) $\ddagger = 0$. In this case, we have only standard levels, which are determined from the condition that $\Gamma\left(-\frac{1}{4}\sqrt{\frac{2m}{g}}E + \frac{1}{2} + \frac{P}{2}\right)$ has poles

$$-\frac{1}{4}\sqrt{\frac{2m}{g}}E + \frac{1}{2} + \frac{P}{2} = -n_r; n_r = 0,1,2... \tag{4.15}$$

ii) $\ddagger = \pm\infty$. In this case, we have only additional levels, which are determined from the condition that $\Gamma\left(-\frac{1}{4}\sqrt{\frac{2m}{g}}E + \frac{1}{2} - \frac{P}{2}\right)$ has poles

$$-\frac{1}{4}\sqrt{\frac{2m}{g}}E + \frac{1}{2} - \frac{P}{2} = -n_r; n_r = 0,1,2... \tag{4.16}$$

Thus, in cases i) and ii) we get analytical expressions of standard and additional levels

$$E_{st,add} = 2\sqrt{\frac{g}{2m}}\{2n_r + 1 \pm P\} \quad n_r = 0,1,2... \tag{4.17}$$

where (+) and (–) signs correspond to standard and additional levels.
iii) For other values of the $\ddagger$ parameter (4.14), the equation is discussed in the appendix.
Let's note that so far in the textbooks [13,14,18] only the formula of standard levels (4.16) with the sign + in front of $P$ was known. So, equation (4.14) is a new, original equation!
We can rewrite the expression (4.17), like the Valence model [10] as follows

$$E_{st,add} = 2\sqrt{\frac{g}{2m}}\left\{n + \frac{3}{2} \pm P - \left(l + \frac{1}{2}\right)\right\} \tag{4.18}$$

Where this time the principal quantum number is defined as follows

$$N = 2n_r + l; \quad n_r = 0,1,2...; l = 0,1,2... \tag{4.19}$$

and for standard levels, let's introduce the concept of quantum defect

$$\Delta_l^{st} = P - \left(l + \frac{1}{2}\right) \qquad (4.20)$$

(4.20) is a physically correct definition. Indeed, if we consider $V_0$ as a small parameter and expand the root of $P$ in the expression (2.8), we get from (4.20)

$$\Delta_l^{st} = -\frac{2mV_0}{2l+1} \qquad (4.21)$$

From which it is clear that for $V_0 = 0$, $\Delta_l^{st} = 0$ i.e. for the potential of (4.1) we have defined a quantum defect as a deviation (defect) from the potential. In general, the quantum defect for the potential $V = -\frac{V_0}{r^2} + W(r)$ can be deined by the formula (4.20) and physically, the quantum defect will be a deviation from the $W(r)$ potential.

For additional levels, it is impossible to carry out the procedure described above, because $V_0$ it is bounded from below according to (2.6).

It is possible to discuss small $V_0$ ones, only for $l = 0$. In this case

$$P = \sqrt{\frac{1}{4} - 2mV_0} \approx \frac{1}{2}(1 - 4mV_0) \qquad (4.22)$$

It can be $V_0$ any small, but it can't be zero, because then $P = 1/2$, and as we mentioned above, we have no additional levels.

Now let us show that for only three values of the τ parameter $ǂ = 0$, $ǂ = \pm\infty$, the property of equidistance of the levels representing the oscillatory potential is maintained {13,14,18}. Indeed, from (4.17) we get the equality confirming this fact

$$E_{st,add}^{n_r+1} - E_{st,add}^{n_r} = 4\check{S}; \quad \check{S} = \sqrt{\frac{g}{2m}} \qquad (4.23)$$

For other values of the τ parameter, this property is violated, to show that, let's denote the energy-dependent left side of equation (4.14) by $f(E)$ and make the ratio $\frac{f(E+4\check{S})}{f(\check{S})}$. It is easy to show that this relation by using the equality

$$\Gamma(z+1) = z\Gamma(z) \qquad (4.24)$$

is equal to following expession

$$\frac{f(E+4\check{S})}{f(E)} = \frac{-\frac{E}{4\check{S}} + \frac{P}{2} - \frac{1}{2}}{-\frac{E}{4\check{S}} - \frac{P}{2} - \frac{1}{2}} \qquad (4.25)$$

From which it is clearly seen that this ratio is not equal to 1 and so the property of equidistance of levels is violated! This fact once again confirms the opinion expressed in [10] that the self-adjusted expansion procedure can change the physical picture.

Note that in equation (4.14) in certain extreme cases we can obtain a clear dependence of the energy on the τ extension parameter. Consider two cases

a) $\ddagger = 0$ is near (close to the standard levels) if we use the well-known formulas for expansion near the poles of Γ functions [27], we get the following formula for the energy

$$E = 2\sqrt{\frac{g}{2m}}\{2(n_r - \nu) + P + 1\}; \quad n_r = 0,1,2... \quad (4.26)$$

where

$$\nu = \frac{(-1)^{n_r}}{n_r!} \frac{\Gamma(1-P)}{\Gamma(1+P)\Gamma(-n_r-1)} \ddagger \ll 1 \quad (4.27)$$

b) $\ddagger = \pm\infty$ near (additional levels) we will have expressions similar to (4.26), but in formulas (4.26) and (4.27) we have to make the following changes $P \to -P$, $\ddagger \to \frac{1}{\ddagger}$.

Now let's write the wave function (4.11) by means of one function, for which we use the following formula from [27]

$$y_5 = \Psi(a,b,z) = \frac{f}{\sin fb}\left\{\frac{F(a,b,z)}{\Gamma(1+a-b)\Gamma(b)} - z^{1-b}\frac{F(1+a-b, 2-b; z)}{\Gamma(a)\Gamma(2-b)}\right\} \quad (4.28)$$

Then from formulas (4.3)-(4.5), (4.11), (4.13) and (4.28) we get the desired result

$$R(r) = Ce^{-\frac{\sqrt{2mg}}{2}r^2}(2mg)^{\frac{1}{4}\left[P-\frac{1}{2}\right]}r^{-\frac{1}{2}+P}\Gamma(1+P)\Gamma(-P-n)\frac{\sin f(1+P)}{f}\Psi(-n, 1+P; \sqrt{2mg}\,r^2) \quad (4.29)$$

Thus (4.29) is a new formula.

We can also represent the $R(r)$ wave function express by Whitaker functions if we use the well-known formula [27]

$$W_{y\sim}(x) = e^{-\frac{x}{2}}x^{\frac{b}{2}}\mathcal{E}(a,b;x); \quad y = -a + \frac{b}{2}; \quad \sim = \frac{b}{2} - \frac{1}{2} \quad (4.30)$$

and we will get

$$R(r) = C(2mg)^{-\frac{3}{8}\left[P-\frac{1}{2}\right]}\Gamma(1+P)\Gamma(-P-n)\frac{\sin f(1+P)}{f}W_{n+\frac{1+P}{2},\frac{P}{2}}(\sqrt{2mg}\,r^2) \quad (4.31)$$

Because the $W_{a,b}(x)$ Whittaker function decays exponentially at infinity [27]

$$W_{a,b}(x) \approx e^{-\frac{1}{2}x}x^a, \quad (4.32)$$
$$x \to \infty$$

It is clear that (4.32) corresponds to the bound states. In addition, (4.31) satisfies the fundamental condition (2.1) in the $0 < P < 1/2$ interval.

Thus, for $\ddagger = 0, \pm\infty$ the standard and additional levels of are given by the formula (4.17), which correspond to the wave functions

$$R_{st} = Ce^{-\frac{\sqrt{2mg}}{2}r^2}(2mg)^{\frac{P-\frac{1}{2}}{4}}r^{\frac{1}{2}+P}F\left(-n,1+P;\sqrt{2mg}\,r^2\right) \quad (4.33)$$

$$R_{add} = De^{-\frac{\sqrt{2mg}}{2}r^2}(2mg)^{\frac{-P-\frac{1}{2}}{4}}r^{\frac{1}{2}-P}F\left(-n-P,1-P;\sqrt{2mg}\,r^2\right) \quad (4.34)$$

The energy of any $\ddagger \neq 0, \pm\infty$ - is determined from the transcendental equation (4.14), and the wave function is given by (4.31). Note that (4.31) is a new formula and we obtained it due to the self-adjoint expansion.

We find the $C$ coefficient (4.31) in the wave function from the normalization probe

$$\int_0^\infty R^2(r)r^2 dr = 1 \quad (4.35)$$

It should be noted that to calculate the (4.35) integral we need the integral of the table [28]

$$\int_0^\infty [W_{y,\sim}(z)]^2 \frac{dz}{z} = \frac{f}{\sin 2fy} \frac{Œ\left(\frac{1}{2}+\sim -y\right)-Œ\left(\frac{1}{2}-\sim -y\right)}{\Gamma\left(\frac{1}{2}+\sim -y\right)\Gamma\left(\frac{1}{2}-\sim -y\right)} \quad (4.36)$$

which convergency condition is

$$|\text{Re}\sim| < \frac{1}{2} \quad (4.37)$$

In our case $y = n + \frac{1+P}{2}; \sim = \frac{P}{2}$, and because $0<P<1/2$, the convergency condition (4.37) is satisfied. From (4.35) and (4.36) we get

$$C^2 = 2f(2mg)^{\frac{3}{4}} \frac{\Gamma(-n)}{\Gamma^2(1+P)\Gamma(-n-P)\sin fP} \frac{1}{[\Psi(-n)-Œ(-n-P)]} \quad (4.38)$$

At the end of this chapter, we make the following remark.

If we consider other solutions (4.9) of equation (4.8)), we have the following situation
1) $y_5$ and $y_7$ the pair has no levels.
2) $y_1$ and $y_5$ the pair only gives standard levels, so it does not give us anything new.
3) $y_2$ and $y_5$ the pair gives us only additional levels ( $\ddagger = \pm\infty$ ), which is a physically unjustified result, since the standard levels are completely lost.
4) $y_2$ and $y_7$ weaning is not allowed, because in this case we do not have standard levels
5) $y_1$ and $y_7$ the pair is inadmissible because the potential $V = -\frac{V_0}{r^2}$ has no levels in the limit $g \to 0$, while in [10,32] we showed that this potential has one negative level.

Thus, only the $y_1$ and $y_2$ pair discussed above is physically interesting.

# V. Conclusion

In this work, it is shown that it is necessary to maintain the second "singular" solution in Schrödinger's equation (2.5), which in turn leads to the necessity of conducting a self-adjoint extension procedure, and it is shown that the parameter of this procedure comes from the requirement of orthogonality.

This general approach is applied to the singular oscillator potential (4.1) and the transcendental equation of eigenvalues (4.14) is obtained. Limiting cases of this equation are studied, and its general investigation is carried out in the appendix. It is shown that the self-adjoint extension violates the equidistant property of energy levels well-known in conventional quantum mechanics for the oscillatory potential, which once again confirms the opinion expressed in [10] that the self-adjoint extension procedure can change the physical picture.

In general, the concept of quantum defect (4.20) is introduced and the wave function of the problem is written as one of the functions (4.29) and (4.31). It is necessary to note that (4.29) and (4.31) are new formulas and are obtained due to the self-adjoint extension procedure.

## Appendix

Let's study the transcendental equation (4.14), the left part of which coincides with the precision of notation [30] with the left side of the labor equation (6.16). In [30], the one-dimensional three-body problem for the oscillator and the inverse square potential in the Schrödinger equation is studied in accordance with the well-known Calogero model [31] and the full self-adjoint extension procedure in the angular and radial Hamiltonians is carried out. In [30] the energy dependence of the left-hand side of equation (6.16) is studied in detail and we can use these results in our equation (4.14). In particular, consider the function

$$f_P(E) = \frac{\Gamma\left(-\frac{E}{4\check{S}} + \frac{1}{2} - \frac{P}{2}\right)}{\Gamma\left(-\frac{E}{4\check{S}} + \frac{1}{2} + \frac{P}{2}\right)} \tag{a.1}$$

as a function of $E$ energy. This function has zeros on the standard levels

$$E_{st} = E_{n_r}^0 = 2\check{S}(2n_r + 1 + P)\,(n_r = 0,1,2...) \tag{a.2}$$

and occurs at $E_{n_r}^\infty = \pm\infty$, which correspond to addition levels

$$E_{add} = E_{n_r}^\infty = 2\check{S}(2n_r + 1 - P)\,(n_r = 0,1,2...) \tag{a.3}$$

Using the results of [30] work, we can show that the function $f_P(E)$ increases monotonically from 0 to $+\infty$, as the energy changes from $E = 0$ to $E_0^\infty$.

Furthermore, since P is in the interval (2.10), we obtain from the (a.2) and (a3)

$$E_{n_r}^\infty < E_{n_r}^0 < E_{n_r+1}^\infty;\ \ \forall_{n_r} = 0,1,2... \tag{a.4}$$

Based on work [30], we can also show that the function $f_P(E)$ increases monotonically in the area (a.4). Moreover, this function is negative in the area $E_{n_r}^\infty < E < E_{n_r}^0$; $\forall_{n_r} = 0,1,2...$ and positive in the area $E_{n_r}^0 < E < E_{n_r+1}^\infty$; $\forall_{n_r} = 0,1,2...$. The right side of equation (4.14) is independent on energy. Therefore, we have the following situation: the physical picture depends on the sign of the ‡ parameter.

1) For ‡ > 0 n the interval $0 < E < E_0^\infty$ we will not have a level, and in all other intervals $[E_{n_r}^\infty, E_{n_r+1}^\infty]$ we have one positive level. Thus, we got that the potential for positive ‡ , (4.1) does not have a negative level at all, which is a physically incorrect result, because the potential (4.1) turns into a $V = -\frac{V_0}{r^2}$ potential in the limit $g \to 0$, which, as we showed in [10,32], has one negative level. Therefore, it can be concluded that the area ‡ > 0 should be excluded.

This is another example of how the "open" ‡ parameter can be limited due to physical requirements. (See [10,32] for other examples of limiting this parameter)

2) ‡ < 0. In this case, the physical picture depends on the value of the following quantity

$$f_P(0) = \frac{\Gamma\left(\frac{1}{2} - \frac{P}{2}\right)}{\Gamma\left(\frac{1}{2} + \frac{P}{2}\right)} \quad (a.5)$$

And we must distinguish two subcases

a) If the condition is fulfilled

$$f_P(0) > -‡ \frac{\Gamma(1-P)}{\Gamma(1+P)} \quad (a.6)$$

Then we will definitely have at least one negative level, and in all other $[E_{n_r}^\infty, E_{n_r+1}^\infty]$ intervals we have one positive level.

b)

$$f_P(0) < -‡ \frac{\Gamma(1-P)}{\Gamma(1+P)} \quad (a.7)$$

Then we have no negative level, we have one positive level in the interval $[0, E_0^\infty]$ and one positive level in all $[E_{n_r}^\infty, E_{n_r+1}^\infty]$ other intervals.

Thus, from the discussion discussed above, it is clear that only subcase a) is physically valid (because it has at least one negative level) and from inequalities (a.5) and (a.7), we get the following restriction on the ‡ parameter

$$‡ > -\frac{\Gamma(1-P)\Gamma\left(\frac{1}{2} + \frac{P}{2}\right)}{\Gamma(1+P)\Gamma\left(\frac{1}{2} - \frac{P}{2}\right)} \quad (a.8)$$